\documentstyle[12pt]{article}
\def\nabstar#1{\nabla\kern-0.5pt\smash{\raise 4.5pt\hbox{$\ast$}}
               \kern-4.5pt_{#1}}

\def\drvstar#1{\partial\kern-0.5pt\smash{\raise 4.5pt\hbox{$\ast$}}
               \kern-5.0pt_{#1}}

\def\newline{\relax\ifhmode\null\hfil\break\else\nonhmodeerr@\newline\fi}
\def\frac#1#2{{#1\over#2}}
\def\text#1{{\hbox{\rm #1}}}
\def\flushpar{{\par \noindent}}

\newcommand{\beq}{\begin{equation}}
\newcommand{\eeq}{\end{equation}}
\newcommand{\bea}{\begin{eqnarray}}
\newcommand{\eea}{\end{eqnarray}}
\def\Id{ \mbox{1\hspace{-1.2mm}I} }
\def\BE{\begin{equation}}
\def\EE{\end{equation}}
\def\BA{\begin{eqnarray}}
\def\EA{\end{eqnarray}}
\def\BAN{\begin{eqnarray*}}
\def\EAN{\end{eqnarray*}}

\def\tr{\mbox{tr}}

\def\det{\mbox{det}}

\def\gm5{\gamma_5}

\def\CT{{\cal T}}

\def\anxc{{\cal A}_D(x)}
\def\Dcont{{\cal D}}
\input epsf.sty
\newdimen\psfigsize
\def\psfigure#1 #2 #3 #4 #5{
    \begin{figure}[tbh]
      \begin{center}
      \vbox{
        \null\vskip-0.2in\hskip#2
        \epsfxsize=#1
        \epsfbox{#4}
        \vskip -0.3in
        \caption {#5 \label{#3}}
        \vskip 0.0 true in plus 0.3 true in
      }
      \end{center}
   \end{figure}
}
\begin{document}
\thispagestyle{empty}
\begin{flushright}
NTUTH-99-121 \\
December 1999
\end{flushright}
\bigskip\bigskip\bigskip
\vskip 2.5truecm
\begin{center}
{\LARGE {Topologically invariant transformation for lattice fermions}}
\end{center}
\vskip 1.0truecm
\centerline{Ting-Wai Chiu}
\vskip5mm
\centerline{Department of Physics, National Taiwan University}
\centerline{Taipei, Taiwan 106, Republic of China.}
\centerline{\it E-mail : twchiu@phys.ntu.edu.tw}
\vskip 2cm
\bigskip \nopagebreak \begin{abstract}
\noindent

A transformation is devised to convert any lattice Dirac fermion
operator into a Ginsparg-Wilson Dirac fermion operator.
For the standard Wilson-Dirac lattice fermion operator,
the transformed new operator is local, free of $ O(a) $ lattice
artifacts, has correct axial anomaly in the trivial sector, and
is not plagued by the notorious problems ( e.g., additive mass
renormalization ) which occur to the standard Wilson-Dirac
lattice fermion operator.


\end{abstract}
\vskip 1.5cm

\newpage\setcounter{page}1

\section{ Introduction }

To formulate chiral fermions on the lattice, one must take into
account the constraints imposed by the Nielson-Ninomiya
no-go theorem \cite{no-go}. It asserts that any lattice
Dirac operator $ D $ in the free fermion limit must violate
at least one of the four basic properties [{\bf (i)-(iv)} listed below ]
of massless Dirac fermion in continuum. Nevertheless, we can incorporate
the gauge interactions in the no-go theorem as follows.
Since the gauge link variables are trivial in the Dirac space,
the chiral symmetry of $ D $ is not affected by turning on
a background gauge field. In other words, if $ \{ D, \gm5 \} $ is zero
in the free fermion limit, then it remains to be zero even when the
gauge links are different from the identity.
Moreover, the presence of gauge interactions does not improve the
locality of $ D $. Therefore, we can assert that any gauge covariant
Dirac operator $ D $ on a finite lattice must violate at least one of the
following properties :
\begin{description}
\item[(i)] $ D $ is chirally symmetric. \newline
           ( $ D \gm5 + \gm5 D = 0 $. )
\item[(ii)] $ D $ is local. \newline
       ( $ | D(x,y) | \sim \exp( - | x - y | / l ) $ with $ l \sim a $; or
         $ D(x,y) = 0 $ for $ |x-y| > z $, where $ z $ is much less than
         the size of the lattice. )
\item[(iii)] In the free fermion limit, $ D $ is free of species doublings.
             \newline
         ( The free fermion propagator $ D^{-1}(p) $ has only one simple
           pole at the origin $ p = 0 $ in the Brillouin zone. )
\item[(iv)] In the free fermion limit, $ D $ has correct continuum behavior.
            \newline
         ( In the limit $ a \to 0 $, $ D(p) \sim i \gamma_\mu p_\mu $
           around $ p = 0 $. )
\end{description}

During the last two years, it has become clear that the proper
way to circumvent the no-go theorem is to break
the chiral symmetry of $ D $ {\bf (i)} at finite lattice spacing,
according to the Ginsparg-Wilson relation \cite{gwr}
\bea
\label{eq:gwr}
D \gm5 ( \Id - a R D ) + ( \Id - a D R ) \gm5 D = 0 \ ,
\eea
where $ R $ is any positive definite Hermitian operator which
is local in the position space and commutes with $\gm5$.
However, the GW relation should be regarded as a generalized
chiral symmetry which contains the usual chiral symmetry
{\bf (i)} in the continuum limit $ a \to 0 $.

Since the massless Dirac operator in continuum is chirally symmetric
( $ \Dcont \gm5 + \gm5 \Dcont = 0 $ ) and antihermitian
( $ \Dcont^{\dagger} = - \Dcont $ ), so it is
$\gm5$-Hermitian ( $ \Dcont^{\dagger} = \gm5 \Dcont \gm5 $ ).
Thus, we require that the lattice Dirac operator $ D $ also preserves
this symmetry at any lattice spacing, i.e.,
\bea
\label{eq:hermit_D}
D^{\dagger} = \gm5 D \gm5 \ .
\eea

The first {\it explicit} and {\it physical} $ D $ satisfying
the GW relation (\ref{eq:gwr}) ( with $ R = 1/2 $ ) and $\gm5$-Hermiticity
is the Overlap Dirac operator
$ D = a^{-1} ( \Id + \gamma_5 \epsilon( H_w ) ) $
which was obtained by Neuberger \cite{hn97:7} in the framework of the
Overlap formalism \cite{rn95,hn99:11}, before the GW
relation was rediscovered. Since the Overlap was essentially motivated
by the Domain-Wall fermion \cite{kaplan92} which was originally
apart from the Ginsparg-Wilson fermion, this seems to indicate that
the Overlap plays a more fundamental role in resolving the problem
of chiral fermions on the lattice than the GW relation.
Thus the Overlap possesses the generalized chiral
symmetry (\ref{eq:gwr}) on the lattice as one of its ingredients.
This in turn suggests that any nonperturbative formulation of chiral
fermions must comply with the exact chiral symmetry,
the GW relation (\ref{eq:gwr}).

One of the salient features of any GW Dirac operator
satisfying {\bf (ii)-(iv) } and (\ref{eq:hermit}) is that
it is not plagued by the notorious problems \cite{ph98:2}
( e.g., additive mass renormalization, etc. )
which usually occur to the Wilson-Dirac
lattice fermion operator. However, if one uses the Overlap Dirac operator
in lattice QCD, one may encounter the technical problem of taking the
square root of a huge positive definite Hermitian matrix $ H_w^2 $,
which would become quite time-consuming when the condition number of
$ H_w^2 $ gets too large. Therefore, both in principle and in practice,
it is important to understand to what extent one can construct
a {\it topologically proper} $ D $ which {\it satisfies
all physical constraints} {\bf (ii)-(iv)}, (\ref{eq:gwr}) and
(\ref{eq:hermit_D}).

The general solution of the Ginsparg-Wilson relation (\ref{eq:gwr})
can be formally written as \cite{twc98:6a,twc98:9a}
\bea
\label{eq:gen_sol}
D = D_c ( \Id + a R D_c )^{-1} = ( \Id + a D_c R )^{-1} D_c
\eea
where $ D_c $ is any chirally symmetric ( $ D_c \gm5 + \gm5 D_c = 0 $ )
Dirac operator which must violate at least one of the three properties
{\bf (ii)-(iv) } listed above.
Now we require $ D_c $ to satisfy {\bf (iii)} and {\bf (iv)},
but violate {\bf (ii)} ( i.e., $ D_c $ is nonlocal ),
since (\ref{eq:gen_sol}) can transform the nonlocal $ D_c $ into a
local $ D $ on a finite lattice for $ R = r \Id $
with $ r $ in the proper range \cite{twc99:10}, and also
preserves the properties {\bf (iii)} and {\bf (iv)}.

Then the $\gm5$-Hermiticity of $ D $ (\ref{eq:hermit_D}) is
equivalent to the $\gamma_5$-Hermiticity of $ D_c $
\bea
\label{eq:hermit}
 D_c^{\dagger} = \gm5 D_c \gm5 \ ,
\eea
due to the relation
\bea
D_c = D ( \Id - a R D )^{-1}  = ( \Id - a D R )^{-1} D \ ,
\eea
which is the inverse transformation of (\ref{eq:gen_sol}).
Then the chiral symmetry and the $\gm5$-Hermiticity of $ D_c $ together
implies that $ D_c $ is antihermitian
\bea
\label{eq:antihermit}
D_c^{\dagger} = -D_c \ ,
\eea
which is in agreement with the antihermiticity of the massless Dirac
operator in continuum. Then there exists one to one correspondence between
$ D_c $ and a unitary operator $ V $ such that
\bea
\label{eq:DcV}
D_c = M a^{-1} (\Id + V )(\Id - V )^{-1}, \\
\label{eq:VDc}
V = (a D_c - M)( a D_c + M)^{-1}.
\eea
where $ M $ is a parameter. Then $ V $ also satisfies the
$\gm5$-Hermiticity, $ V^{\dagger} = \gamma_5 V \gamma_5 $.
Substituting (\ref{eq:DcV}) into (\ref{eq:gen_sol}),
we obtain \cite{twc98:6a}
\bea
\label{eq:gensolV}
D = M a^{-1} ( \Id + V )[ ( \Id - V ) + M R ( \Id + V ) ]^{-1} \ .
\eea
Note that $ D $ is well-defined even if $ D_c $ has singularities
in nontrivial gauge backgrounds. In this case, the index of $ D_c $
is defined to be the index of $ ( \Id + V ) $, the numerator on the
r.h.s. of (\ref{eq:DcV}). Evidently, the zero modes and the index of
$ D $ are invariant for any $ R $ in (\ref{eq:gen_sol})
or (\ref{eq:gensolV}) \cite{twc98:6a,twc98:9a}.
That is, a zero mode of $ D_c $ is also a zero mode of $ D $
and vice versa, hence,
\beq
\label{eq:npm}
 n_{+} ( D_c ) = n_{+} ( D ), \hspace{4mm} n_{-} ( D_c ) = n_{-} ( D ),
\eeq
\beq
\label{eq:index}
\mbox{index}(D_c) = n_{-}(D_c) - n_{+}(D_c) =
 n_{-}(D) - n_{+}(D) = \mbox{index}(D) \ ,
\eeq
where $ n_{+} ( n_{-} ) $ denotes the number of zero modes of
$ +1 ( -1 ) $ chirality.

The conditions {\bf (ii)-(iv)}, (\ref{eq:gwr}) and (\ref{eq:hermit_D})
constitute the neccesary conditions for $ D $ to reproduce the continuum
axial anomaly on a finite lattice. For the trivial sector, they are
sufficient to guarantee that the correct axial anomaly can be
recovered on the lattice.

However, for topologically nontrivial gauge backgrounds,
the index as well as the axial anomaly of a lattice Dirac operator $ D $
depends on its topological characteristics \cite{twc99:11}.
If $ D $ is topologically proper ( i.e., satisfying
the Atiyah-Singer index theorem for any gauge background
satisfying the topological bound \cite{twc99:11} ),
then the sum of the axial anomaly of $ D $ over
all sites on a finite lattice is equal to the topological charge
of the gauge background.
Then it follows that the axial anomaly of $ D $ would agree with
the topological charge density of the gauge background
if $ D $ is local.

Now the central problem is how to construct a chirally symmetric
and nonlocal $ D_c $ which satisfies {\bf (i)}, {\bf (iii)},
{\bf (iv)}, and (\ref{eq:hermit}). Furthermore we also require
that $ D_c $ is topologically proper.
These constitute the necessary requirements
\cite{twc98:9a} for $ D_c $ to enter (\ref{eq:gen_sol})
such that $ D $ could provide a nonperturbative regularization
for a massless Dirac fermion interacting with a background gauge field.
A systematic construction of $ D_c $ satisfying {\bf (i)}, {\bf (iii)},
{\bf (iv)}, and (\ref{eq:hermit}) has been discussed in ref. \cite{twc99:8}.
However, a general prescription for constructing a topologically
proper $ D_c $ ( $ D $ ) still remains an unsolved problem.

Nevertheless, given any lattice Dirac operator $ D $,
one can always transform $ D $ into a GW Dirac operator
$ D' $ which preserves all essential properties of $ D $.
In other words, if $ D $ is topologically proper and satisfies
{\bf (ii)-(iv)} and ({\ref{eq:hermit_D}), then the
GW Dirac operator $ D' $ also possesses these properties.

The outline of this paper is as follows. In Section 2, we
define the topologically invariant transformations \cite{twc99:6}
for lattice Dirac operators. The transformation which can convert
any lattice Dirac operator into a GW Dirac operator is
derived. In Section 3, we apply this transformation
to the Wilson-Dirac lattice fermion operator and
obtain a new GW Dirac operator which is local, free of $ O(a) $
lattice artifacts and has correct axial anomaly in the trivial sector.
In Section 4, we discuss and summarize.

\section{ Topologically invariant transformation }

Given any lattice Dirac operator $ D $, in general, there are many
different ways to extract its chirally symmetric part.
For example, one can construct
\bea
\label{eq:Ds}
D_s = \frac{1}{2} ( D - \gamma_5 D \gamma_5 )
\eea
which is chirally symmetric ( $ D_s \gamma_5 + \gamma_5 D_s = 0 $ ).
However, (\ref{eq:Ds}) does not necessarily preserve the
property {\bf (iii)}. For example, if one applies this transformation
to the Wilson-Dirac operator $ D_W = \gamma_\mu t_\mu + W $
(\ref{eq:DW}), one obtains $ D_s = \gamma_\mu t_\mu $, the naive
fermion operator which suffers from the species doublings.
Although $ D_W $ is free of species doublings in the continuum limit,
the transformation (\ref{eq:Ds}) cannot preserve this property
since $ D_s $ satisfies {\bf (i), (ii)} and {\bf (iv)},
thus it must violate {\bf (iii)} as a consequence of the no-go theorem.
Therefore, we need a transformaton which preserves
the properties {\bf (iii)}, {\bf (iv)}, (\ref{eq:hermit_D})
and (\ref{eq:npm}), but exchanges the locality of $ D $ for
its chiral symmetry at finite lattice spacing.

Consider the topologically invariant
transformation \cite{twc99:6} on any $\gm5$-Hermitian $ D $,
\bea
\label{eq:tit}
\CT(R) : \hspace{4mm} D \to D'= \CT(R)[D] \equiv D ( \Id + a R D )^{-1}
                               = ( \Id + a D R )^{-1} D
\eea
where $ R $ is a Hermitian operator which commutes
with $ \gm5 $ and satisfies \\
$ \det( \Id + a R D ) \ne 0 $. Then $ D' $ is also
$\gm5$-Hermitian.

It is obvious that a zero mode of $ D $ is a zero mode of $ D' $ and
vice versa. Suppose $ \phi $ is a zero mode of $ D $, i.e., $ D \phi = 0 $,
then $ D' \phi = ( \Id + a D R )^{-1} D \phi = 0 $. On the other
hand, if $ \phi $ is a zero mode of $ D' $, i.e., $ D' \phi = 0 $,
then we can use the relation $ ( \Id + a D R ) D ' = D $ to obtain
$ D \phi = 0 $. Therefore
\BAN
          D \phi = 0 \Leftrightarrow  D' \phi = 0 \ .
\EAN
Moreover, if the zero modes of $ D $ have definite chirality, i.e.,
$ D \phi = 0 $ implies $ D \gm5 \phi = 0 $, then the chirality of
$ \phi $ is preserved under the transformation (\ref{eq:tit}), i.e.,
$ D' \gm5 \phi = 0 $. In this case, we have
\BAN
 D \phi_{\pm} = 0 &\Leftrightarrow&  D' \phi_{\pm} = 0 \ , \\
    n_{\pm} ( D') &=& n_{\pm} (D)  \ ,     \\
 \mbox{index}(D') &=& \mbox{index}(D) \ .
\EAN

In the trivial sector, $ D $ is nonsingular
except possibly some "exceptional" configurations of
zero measure, thus (\ref{eq:tit}) is equivalent to
\bea
\label{eq:titi}
\CT(R) : \hspace{4mm} D \to D' \mbox{ such that } D'^{-1} = D^{-1} + a R \ .
\eea

Formally, there exists a set of transformations $ \{ \CT(R) \} $
which form an abelian group with group multiplication
\bea
\label{eq:mult}
 \CT(R_1) \circ \CT(R_2) \ [D] = \CT(R_1) \left[ \ \CT(R_2)[D] \ \right]
                               = \CT( R_1 + R_2 ) [D] \ ,
\eea
where the last equality follows immediately from (\ref{eq:titi}).
Explicitly, one checks :\\
(a) the closure property is satisfied since the sum of any two
    ($\gm5$-)Hermitian operators is ($\gm5$-)Hermitian;        \\
(b) the identity element is $ \CT(0) $; \\
(c) the inverse of $ \CT(R) $ is $ \CT(-R) $; \\
(d) the associative law
\BAN
& & ( \CT(R_1) \circ \CT(R_2) ) \circ \CT(R_3) =
    \CT(R_1) \circ ( \CT(R_2) \circ \CT(R_3) ) \\
&=& \CT( R_1 + R_2 + R_3 )
\EAN
is also satisfied. \\

The chiral limit of the transformation (\ref{eq:tit}) is at
\bea
\label{eq:Rc}
R = - \frac{1}{2} a^{-1} ( D^{-1} + \gamma_5 D^{-1} \gamma_5 ) \equiv R_c \
\eea
which gives the chirally symmetric $ D_c $
\bea
\label{eq:D5D}
D_c = \CT(R_c) [D] = 2 \gm5 D ( \gm5 D - D \gm5 )^{-1} D \ .
\eea

Substituting $ D_c $ (\ref{eq:D5D}) into (\ref{eq:gen_sol}),
we obtain a GW Dirac operator
\bea
\label{eq:GW_D}
D' = 2 \gm5 D ( \gm5 D - D \gm5 + 2 a D R \gm5 D )^{-1} D \ ,
\eea
which satisfies the GW relation
\BAN
  D' \gm5 + \gm5 D' = 2 a D' R \gm5 D' \ ,
\EAN
where $ R $ is a positive definite Hermitian operator which is
local in the position space, commutes with $ \gm5 $,
and is chosen such that the inverse operator on the
r.h.s. of (\ref{eq:GW_D}) is well-defined.
In the following, we shall denote the GW Dirac operator
(\ref{eq:GW_D}) by
\BAN
D' = \CT( R ) [D_c] = \CT(R) \circ \CT( R_c ) [D] = \CT( R + R_c ) [D] \ .
\EAN

Note that if $ D $ has zero modes in topologically nontrivial
gauge backgrounds, then $ D_c $ (\ref{eq:D5D}) has singularities
and is not well-defined, however, the GW Dirac operator $ D' $
(\ref{eq:GW_D}) is still well-defined. For example, consider the
$\gm5$-hermitian Dirac operator $ D = a^{-1} ( \Id + V ) $, where $ V $
is a unitary operator having real eigenvalues $ \pm 1 $ in nontrivial
gauge backgrounds. Then it can be shown that the real ( $ \pm 1 $ )
eigenmodes of $ V $ have definite chirality and satisfy the chirality
sum rule, while each complex eigenmode has zero chirality. Thus
each $ +1 $ eigenmode must be accompanied by a $ -1 $ eigenmode
of opposite chirality, and vice versa.
Now Eq. (\ref{eq:D5D}) gives
\BAN
D_c = 2 a^{-1} \frac{ \Id + V }{ \Id - V } \ ,
\EAN
which has a pole if $ D $ has a zero mode.
However, Eq. (\ref{eq:GW_D}) gives a well-defined GW Dirac operator
\BAN
D' = 2 a^{-1} ( \Id + V ) [ ( \Id - V ) + 2 R ( \Id + V ) ]^{-1} \ .
\EAN


It is evident that for any two lattice Dirac operators $ D^{(1)} $ and
$ D^{(2)} $ satisfying (\ref{eq:hermit_D}),
their corresponding chiral limits obtained from (\ref{eq:D5D}),
say, $ D_c^{(1)} $ and $ D_c^{(2)} $, are in general different.
However, they are related by the transformation
\bea
\label{eq:DcT}
D_c^{(1)} = \sum_{i} T_i D_c^{(2)} T_{i}^{\dagger}
\eea
where each $ T_i $ commutes with $ \gm5 $. In general,
given any two $ D_c^{(1)} $ and $ D_c^{(2)} $,
it is nontrivial to obtain all $ T_i $ in (\ref{eq:DcT}).

\section{ Chirally invariant Wilson-Dirac operator }

The Wilson-Dirac lattice fermion operator \cite{wilson75}
can be written as
\bea
\label{eq:DW}
D_W = \gamma_\mu t_\mu + W
\eea
where
\bea
\label{eq:tmu}
t_\mu (x,y) = \frac{1}{2a} [   U_{\mu}(x) \delta_{x+\hat\mu,y}
                       - U_{\mu}^{\dagger}(y) \delta_{x-\hat\mu,y} ] \ ,
\eea
\bea
\gamma_\mu &=& \left( \begin{array}{cc}
                            0                &  \sigma_\mu    \\
                    \sigma_\mu^{\dagger}     &       0
                    \end{array}  \right)  \ ,
\eea
\beq
\sigma_\mu \sigma_\nu^{\dagger} + \sigma_{\nu} \sigma_\mu^{\dagger} =
2 \delta_{\mu \nu} \ ,
\eeq
and $ W $ is the Wilson term
\bea
\label{eq:wilson}
W(x,y) =  \frac{1}{2a} \sum_\mu \left[ 2 \delta_{x,y}
                  - U_{\mu}(x) \delta_{x+\hat\mu,y}
                  - U_{\mu}^{\dagger}(y) \delta_{x-\hat\mu,y} \right] \ .
\eea
The color and Dirac indices have been suppressed in (\ref{eq:DW}).
The first term on the r.h.s. of (\ref{eq:DW}) is the naive fermion
operator which satisfies properties {\bf (i), (ii)} and {\bf (iv)} but
violates {\bf (iii)} since it has $ 2^{d} - 1 $ fermion doubled modes.
The purpose of the Wilson term is to give each doubled mode
a mass of $ \sim 1/a $ such that in the continuum limit
each doubled mode becomes infinitely heavy and decouples from the fermion
propagator. However, the introduction of the Wilson term has serious
drawbacks. It causes $ O(a) $ artifacts and also leads to the notorious
problems such as additive fermion mass renormalization,
vector current renormalization, and mixings between operators in
different chiral representations.

The free fermion propagator of $ D_W $ in momentum space is
\bea
\label{eq:Dwi}
D_W^{-1} (p) = ( \gamma_\mu t_\mu )^{-1} \frac{ t^2 }{ w^2 + t^2 }
             + \frac{ w }{ w^2 + t^2 }
\eea
where $ t_\mu = i a^{-1} \sin( p_\mu a ) $,
$ t^2 = a^{-2} \sum_{\mu} \sin^2 ( p_\mu a ) $
and $ w = 2 a^{-1} \sum_{\mu} \sin^2( p_\mu a/2 ) $. We note that
in the first term of (\ref{eq:Dwi}), the doubled modes are decoupled
due to the vanishing of the factor
$ t^2/( w^2 + t^2 ) $ at the $ 2^d - 1 $ corners of the
Brillouin zone. However, the second term in (\ref{eq:Dwi})
breaks the chiral symmetry explicitly, and is the source of additive
mass renormalization and other notorious problems. Any satisfactory
solution to all these problems must get rid of the second term entirely,
while keeping the first term intact.

Applying the topologically invariant transformation (\ref{eq:D5D})
to $ D_W $, we obtain the chirally symmetric $ D_c $,
\bea
\label{eq:DWc}
D_c = \gamma_\mu t_\mu - W \ ( \gamma_\mu t_\mu )^{-1} \ W
\eea
which is antihermitian, nonlocal ( due to the second term ),
free of $ O(a) $ lattice artifacts, and
satisfies {\bf (iii), (iv)} and (\ref{eq:hermit}).

The free fermion propagator of (\ref{eq:DWc}) in momentum space is
\bea
\label{eq:Dci}
D_c^{-1} (p) = ( \gamma_\mu t_\mu )^{-1} \frac{ t^2 }{ w^2 + t^2 }
\eea
which is exactly the first term of (\ref{eq:Dwi}).
So, we have "chiraled away" the second term of (\ref{eq:Dwi}) through
the transformation (\ref{eq:D5D}).

Substituting (\ref{eq:DWc}) into (\ref{eq:gen_sol})
with $ R = r \Id $, we obtain a GW Dirac operator
\bea
\label{eq:DW_GW}
D = \left[ \begin{array}{cc}
 a \ r \ C^{\dagger} C ( \Id  + a^2 r^2 C^{\dagger} C )^{-1}  &
    - C^{\dagger} ( \Id + a^2 r^2 C C^{\dagger} )^{-1}      \\
 C ( \Id + a^2 r^2 C^{\dagger} C )^{-1}  &
 a \ r \ C C^{\dagger} ( \Id + a^2 r^2 C C^{\dagger} )^{-1}
            \end{array}      \right]
\eea
where
\bea
\label{eq:C}
C = ( \sigma^{\dagger}_\mu t_\mu ) - W ( \sigma_\mu t_\mu )^{-1} W \ .
\eea
The locality of $ D $ depends on the gauge configuration as well
as the value of $ r $. If $ r $ is zero, then $ D $ is equal to
$ D_c $ which is nonlocal, since $ D_c $ is chirally symmetric,
free of species doublings and has correct continuum behavior.
For sufficiently smooth gauge configurations, $ D $ is local
( and not highly peaked in diagonal elements ) for $ r $ within
a proper range. For example, consider the gauge configuration
in Fig. 1, $ D $ is local for $ r \in ( 0.2, 0.8 ) $.
A rigorous proof of the locality of $ D $ for a given $ r $ and for
gauge configurations satisfying a certain bound is beyond
the scope of the present paper.

The axial anomaly of $ D $ can be written as
\bea
\label{eq:anxc}
\anxc = \tr [   ( \Id + a^2 r^2 C C^{\dagger} )^{-1}
              - ( \Id + a^2 r^2 C^{\dagger} C )^{-1} ] (x,x)
\eea
where $ \tr $ denotes the trace over the color and spinor space.
Since $ D_c $ (\ref{eq:DWc}) in the free fermion limit is free of
species doubling and has the correct continuum behavior,
the perturbation calculation in ref. \cite{twc99:1} showed that
$ D $ (\ref{eq:DW_GW})
has the correct axial anomaly in the trivial sector.
This has been verified explicitly on finite lattices.
An example is shown in Fig. 1, in which the axial anomaly $ \anxc $
is plotted for each site on a $ 12 \times 12 $ lattice with lattice
spacing $ a = 1 $, comparing with the topological charge density
$ \rho(x) = \frac{1}{2\pi} F_{12}(x) $ of the trivial gauge background.
The position of a site with coordinates $ ( x_1, x_2 ) $ is represented
by an integer $ x = 12 ( x_2 - 1 ) + x_1 $, as the $x$-coordinate in
Fig. 1. The axial anomaly $ \anxc $ is denoted by diamonds, while the
topological charge density $ \rho(x) $ of the gauge background by circles.
The line segments between circles are inserted only for the visual purpose.
Evidently $ \anxc $ agress with the topological charge density $ \rho(x) $
at each site.

It is instructive to compare the axial anomaly of the GW Dirac operator
$ D $ (\ref{eq:DW_GW}) in a trivial gauge background,
as shown in Fig. 1, to that of the Wilson-Dirac
operator $ D_W $ (\ref{eq:DW}) in the same gauge background,
as shown in Fig. 1 in ref. \cite{twc99:11}.
The former agrees with the topological charge density $ \rho(x) $
while the latter does not.
This demonstrates that the transformation $ \CT( r + R_c ) $
indeed plays an important role in converting $ D_W $ into a GW Dirac
operator $ D = \CT( r + R_c ) [ D_W ] $ which is free of $ O(a) $
lattice artifacts. Thus $ D $ can reproduce the correct axial anomaly
even on a {\it finite} lattice.

\psfigure 5.0in -0.2in {fig:anxc} {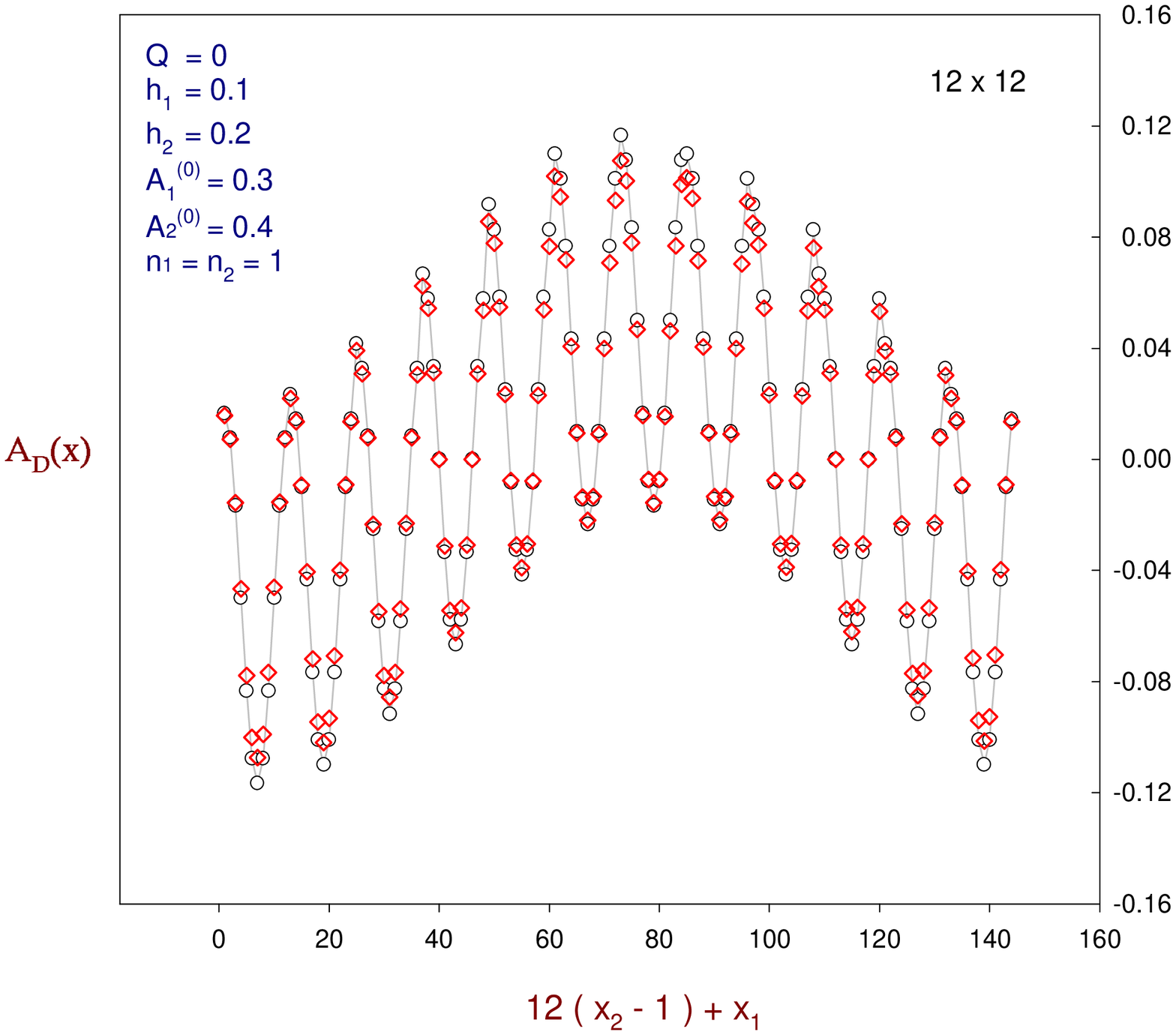} {
The axial anomaly $ \anxc $ [ Eq. (\ref{eq:anxc}) ]
of the massless GW Dirac operator $ D $ [ Eq. (\ref{eq:DW_GW}) ]
in a trivial gauge background on a $ 12 \times 12 $ lattice.
The value of $ r $ has been set to $ 0.5 $, and there is no significant
changes to $ \anxc $ for $ r \in ( 0.2, 0.8 ) $.
The background gauge field is the same as that in Fig. 1 of
ref. \cite{twc99:11}.
The axial anomaly $ \anxc $ is denoted by diamonds.
The topological charge density $ \frac{1}{2 \pi} F_{12} $
of the gauge background is plotted as circles, which are joined by
line segments for the visual purpose.}

However, this GW Dirac operator (\ref{eq:DW_GW}) does not possess
exact zero modes in any topologically nontrivial gauge background
( i.e., $ D $ is topologically trivial ).
This is expected since the Wilson-Dirac operator $ D_W $ is
topologically trivial, hence the number of zero modes
remains zero ( $ n_\pm ( D ) = n_\pm ( D_W ) = 0 $ )
under the transformation $ D_W \to D = \CT( r + R_c )[ D_W ] $.
Consequently, the sum of the axial anomaly over all sites is zero,
not equal to the topological charge of the gauge background.
Therefore the axial anomaly of $ D $ ( or $ D_W $ )
cannot agree with the topological charge density of the
nontrivial gauge background on {\it any} finite lattice.
It follows that {\it the disagreement must persist even in the
continuum limit $ a \to 0 $ } \cite{twc99:11}.

Even though the new GW Dirac operator (\ref{eq:DW_GW}) is
topologically trivial, it has well-defined chiral properties
and is not plaqued by the notorious problems which occur to
the standard Wilson-Dirac operator (\ref{eq:DW}).
Therefore, from this viewpoint,
it seems that (\ref{eq:DW_GW}) may be an alternative
to the Wilson-Dirac operator in lattice QCD,
especially for studies involving observables which are dominated by
the trivial sector. Further implementations of (\ref{eq:DW_GW}) for
lattice QCD are beyond the scope of this paper.

For massive Dirac fermion operator, the mass term $ m $ should enter
$ D $ according to the following transformation \cite{twc98:10b}
\bea
\label{eq:D_m}
\CT_m(R) : \hspace{4mm} D_c \to D = ( D_c + m ) ( \Id + a R D_c )^{-1} \ .
\eea
Then for the massive Wilson-Dirac fermion operator, the transformed GW
Dirac operator with $ R = r \Id $ is
\bea
\label{eq:DW_GW_m}
D = \left[ \begin{array}{cc}
 ( m \Id + a r \ C^{\dagger} C ) ( \Id  + a^2 r^2 C^{\dagger} C )^{-1}  &
    - ( 1 - m \ a \ r ) C^{\dagger} ( \Id + a^2 r^2 C C^{\dagger} )^{-1}      \\
 ( 1 - m \ a \ r ) C ( \Id + a^2 r^2 C^{\dagger} C )^{-1}  &
 ( m \Id + a r \ C C^{\dagger} ) ( \Id + a^2 r^2 C C^{\dagger} )^{-1}
            \end{array}      \right]
\eea
where $ C $ is defined in (\ref{eq:C}).


Although the Wilson-Dirac fermion operator $ D_W $ is topologically
trivial, it can be used to construct the Domain-Wall fermion operator
\cite{kaplan92} on a five dimensional lattice with fermion
mass as a step function in the fifth dimension such that chiral
fermions can be realized on the four dimensional walls where the mass
defects locate. Then the current flowing into the fifth dimension can
induce the correct axial anomaly on the four dimensional walls.
Therefore exact zero modes can be reproduced for the chiral fermions
on the four dimensional walls for topologically nontrivial gauge
backgrounds, and the Atiyah-Singer index theorem can be satisfied.
However, the chiral symmetry of the fermions residing on the
four-dimensional walls is an exact symmetry {\it only} in the limit
the number of sites in the fifth dimension $ N_s $ goes to infinity.
Thus, at finite $ N_s $, the Domain-Wall fermion
may suffer from anomalous effects due to chiral symmetry violations
\cite{wu99}.
It seems that one may use the new GW Dirac operators (\ref{eq:DW_GW})
and (\ref{eq:DW_GW_m}) to construct a corresponding Domain-Wall fermion
operator. Then it would possess the exact chiral symmetry (\ref{eq:gwr})
even at finite $ N_s $, and the absence of additive quark mass
renormalization can be guaranteed.

\section{ Summary and discussions }

We can understand the topologically invariant transformation
$ \CT( r + R_c ) $ by the following considerations.

Given any lattice Dirac operator $ D $, there exists
an operator $ R $ such that the Ginsparg-Wilson
relation (\ref{eq:gwr}) can be satisfied.
Then, according to (\ref{eq:gen_sol}), there exists a chirally
symmetric $ D_c $ such that $ D $ can be written in the form
\bea
\label{eq:gen}
D = D_c ( \Id + a R D_c )^{-1} = \CT(R) [ D_c ] \ .
\eea
Thus the inverse transformation of (\ref{eq:gen}) is
\bea
\label{eq:inverse}
D_c = \CT(-R) [ D ] =  D ( \Id - a R D )^{-1} \equiv \CT(R_c) [ D ]
\eea
where
\bea
\label{eq:R}
R_c = -R = - \frac{1}{2} a^{-1} ( D^{-1} + \gamma_5 D^{-1} \gamma_5 ) \ .
\eea
Once $ D_c $ is obtained, it can be substituted into (\ref{eq:gen_sol})
with $ R $ independent of $ D $. In particular, for
$ R = r \Id $, it gives
\bea
\label{eq:titr}
D' = \CT(r)[ D_c ] = \CT(r) \circ \CT(R_c) [D] = \CT( r + R_c ) [D] \ .
\eea

If $ D $ is topologically proper and
satisfies {\bf (iii)}, {\bf (iv)} and (\ref{eq:hermit_D}),
then these properties are preserved under the transformation
(\ref{eq:titr}), i.e., $ D' $ is topologically proper
and satisfies {\bf (iii)-(iv)}, (\ref{eq:gwr}) and (\ref{eq:hermit_D}).
Then it follows that the axial anomaly of $ D'$ would agree with the
topological charge density of the gauge background, provided that
$ D' $ is local ( with $ r $ in the proper range ).
On the other hand, if $ D $ is topologically trivial, and
satisfies {\bf (iii)-(iv)} and (\ref{eq:hermit_D}),
then the axial anomaly of a local $ D'$ ( with $ r $ in the proper range )
would agree with the topological charge density for a trivial gauge
background, but not for the nontrivial ones.

For the Wilson-Dirac lattice fermion operator (\ref{eq:DW}),
the transformed GW Dirac operator (\ref{eq:DW_GW}) is local,
free of $ O(a) $ lattice artifacts,
has correct axial anomaly in the trivial sector, and
is not plagued by the notorious problems
( e.g., additive mass renormalization, etc. )
which occur to the Wilson-Dirac operator.
It seems that (\ref{eq:DW_GW}) or (\ref{eq:DW_GW_m}) may be an
alternative to the Wilson-Dirac operator in lattice QCD.


\bigskip
\bigskip


\flushpar
{\bf Acknowledgement }
\bigskip

\noindent
This work was supported by the National Science Council, R.O.C.
under the grant number NSC89-2112-M002-017.

\bigskip
\bigskip

\vfill\eject

\end{document}